\documentclass[times, 10pt,twocolumn]{article}
\usepackage{latex8}
\usepackage{times}
\usepackage{amsmath}
\usepackage{graphicx}
\usepackage{amsthm}
\usepackage{subfigure}

\newenvironment{mytinylisting}
{\begin{list}{}{\setlength{\leftmargin}{0em}}\item\tiny\bfseries}
{\end{list}}

\begin{document}

\title{Autoregressive Time Series Forecasting of Computational Demand}

\author{Thomas Sandholm\\
KTH -- Royal Institute of Technology\\
School of Information and Communication Technology\\
SE-16440 Kista, Sweden \\ sandholm@kth.se}

\maketitle
\thispagestyle{empty}

\begin{abstract}
We study the predictive power of 
autoregressive moving average models
when forecasting demand in 
two shared computational networks,
PlanetLab and Tycoon.
Demand in these networks 
is very volatile, and predictive techniques to plan 
usage in advance can improve the performance
obtained drastically. 

Our key finding 
is that a random walk predictor 
performs best for one-step-ahead 
forecasts, whereas ARIMA(1,1,0) and
adaptive exponential smoothing 
models perform
better for two and three-step-ahead 
forecasts. A Monte Carlo
bootstrap test is proposed to evaluate 
the continuous prediction performance
of different models with arbitrary
confidence and statistical
significance levels. 
Although the prediction results differ
between the Tycoon and PlanetLab
networks, we observe very similar 
overall statistical properties, such
as volatility dynamics.
\end{abstract}

\section{Introduction}
Shared computational resources are gaining popularity
as a result of innovations in network connectivity,
distributed security, virtualization and standard
communication protocols. The vision is to use 
computational power in the same way as electrical
power in the future, i.e. as a utility. 
The main obstacle for delivering on that vision is
reliable and predictable performance. Demand can
be very bursty and random, which makes it hard to
plan usage to optimize future performance.
Forecasting methods for aiding usage planning
are therefore of paramount importance for offering
reliable service in these networks.

In this paper we study the demand dynamics of two
time series from computational markets, PlanetLab\footnote{http://www.planet-lab.org}
and Tycoon\footnote{http://tycoon.hpl.hp.com}. 
Our main objective is to study the 
prediction abilities and limitations of time series regression
techniques when forecasting averages
over different time periods. Here we focus on hourly
forecasts that could be applied for scheduling
jobs with run times in the order of a few hours,
which is a very common scenario in these systems. The main
motivation for this study was that an exponential 
smoothing technique used in previous work~\cite{sandholm2007a},
was found to perform unreliably in a live deployment.

The general evaluation approach is to model the structure
of a small sample of the available time series, and assume the
structure is fixed over the sample set. Then
perform predictions with regularly updated model parameters
and benchmark those predictions against a simple
strategy using the current value as the 
one-step-ahead forecast (assuming a random walk).

We focus our study on the following questions.
\begin{itemize}
\item{Can a regression model
perform better 
than a strategy assuming a random walk with no correlations 
in the distant past?}
\item{How much data into the past are
needed to perform optimal forecasts?}
\item{How often do we need to 
update the model parameters?}
\end{itemize}
The answers to these questions depend on both the size of the sliding window
used for the forecast and on the length of the forecast horizon. Our goal
is to give general guidelines as to how forecasts should be performed in this
environment. 
 
When predicting demand in computational
networks
instantaneous, adaptive, flexible,
and light-weight predictors are required
to accurately estimate the risk of service
degradation and to quickly take preemptive
actions. 
With the increased popularity of 
virtualized computational markets such as Tycoon, 
this need for prediction takes a new dimension.
Successful forecasts can now reduce the
cost of computations more directly and explicitly.
However, high volatility and
non-stationarity of demand complicates
model building and reduces prediction
reliability.

The main objective of this study is 
to investigate which time series models
can be used when predicting demand 
in computational markets, and how they 
compare in terms of predictive accuracy
to simpler random walk and exponential 
smoothing models. Since modeling and 
parameter estimations need to 
adapt quickly to regime shifts, a simple
fixed static model of the entire 
series is not likely to produce any good results.
In this work we make a compromise and fix the 
structure of the model but update the parameter
estimates continuously. 

The contribution of this work is threefold:
\begin{itemize}
\item we perform ARIMA modeling and prediction of Tycoon and PlanetLab demand,

about predictor model performance,
\item and we identify common statistical properties of PlanetLab
and Tycoon demand.
\end{itemize}

The paper is structured as follows. In Section~\ref{sec:method} our evaluation
approach is discussed, and in Section~\ref{sec:pl} we model and
predict the PlanetLab series. In Section~\ref{sec:tycoon} we perform the 
same analysis for the Tycoon series. Then we compare the analyses in
Section~\ref{sec:compare} and  discuss related work in Section~\ref{sec:related_work} before concluding
in Section~\ref{sec:conclusion}.

\section{Evaluation Method}\label{sec:method}
In this section, we describe the method used
to construct models and to evaluate the forecasting 
performance of models of the time series
studied.

\subsection{Modeling}
We first construct an autoregressive
integrated moving average (ARIMA) model of a small sample of the
time series in order to determine the general 
regression structure of the data. The rationale
behind this approach is that the majority of the data should be
used to evaluate the forecasting performance. During
forecasting the model parameters are refit, and to compensate for 
possible changes in structure we evaluate a number of 
similar benchmark models. Furthermore, in a real deployment, we
ideally want to re-evaluate the regression structure infrequently
compared to the number of times the structure can be used for 
predictions to make it viable. The sample used for determining 
the regression structure is discarded in the forecasting evaluation
to keep the predictions unbiased. Conversely, no measured
properties of the time series outside of the sample window are used
when building the models of the predictors.

The general
model and the benchmark models are then fit to partitions of the
data in subsequent time windows.
In each time window the model parameters are re-evaluated.
The fitted model then produces one, two, and three-step-ahead 
forecasts. The forecasts are thus conditioned on
the assumption of a specific structure of the model.
The size of the time windows are made small enough to allow
a large number of partitions and thus also independent
predictions, and kept big enough for the 
ARIMA maximum likelihood fits to converge. 

\subsection{Forecast}
The fitted ARIMA model structure is compared to two
standard specialized ARIMA processes. The first
benchmark model used is the random walk model (RW), 
ARIMA(0,1,0), which always produces
the last observed value as the forecast. The second
model is the Exponentially Weighted Moving Average (EWMA), 
a.k.a. the exponential smoothing model, which 
can be represented as an ARIMA(0,1,1) or
IMA(1,1) process producing forecasts with an exponential decay
of contributions from values in the past. This representation
is due to Box et al.~\cite{box1994} who showed that
the optimal one-step-ahead forecast of the IMA(1,1) model
with parameter $\theta$ is the same as the exponential
smoothing value with factor $\lambda=1-\theta$. 

For each set of time-window predictions performed, the
mean square error (MSE) is computed. To facilitate
comparison, the MSEs are normalized against the random walk
model as follows
\begin{equation}
\hat \epsilon = ln(e_m/e_b)
\end{equation}
where $e_m$ is the MSE of the model studied, and $e_b$ is the 
MSE of the benchmark. Thus an $\hat \epsilon > 0$ means
that the model generated more accurate forecasts than the benchmark.
Hence, we have
\begin{equation}\label{model_compare}
F_{m,b}=Pr(e_m \leq e_b)=\int_{-\infty}^0{f_{\hat \epsilon}}
\end{equation}
where $f_{\hat \epsilon}$ is the probability density function (PDF)
of $\hat \epsilon$. 
Thus we have constructed a statistic for evaluating
the models based on the cumulative distribution function (CDF) of
the log ratio of the model and the RW benchmark MSEs, which we call
{\it normalized distribution error} or NDE.
This statistic is similar in spirit to the MSE measurement itself, but
to avoid a bias towards symmetric error distributions, we base
our statistic on the median as opposed to the mean. One might
argue that highly incorrect predictions, therefore, are not penalized
strongly enough, but we are more interested in the reliability
aspect of predictions here, i.e., which model can be trusted to
perform better in most cases. If the error distribution has many 
outliers it should be reflected in the width of the confidence
bound instead. We thus focus next on building such unbiased 
confidence bounds.

\subsection{Statistical Test}
With the NDE statistic we have a metric to decide when a model performs
better than a benchmark, but in order to render claims of
statistical significance and prediction confidence bounds,
a measure of error variance is needed.
Due to a limited set of original data points (one MSE for
each sample window size),
the approach is to use bootstrap sampling based on the 
empirical distribution of $\hat \epsilon$.
Using~(\ref{model_compare}) the null hypothesis is $H_0: F_{m,b} > .5$,
that is, the studied model predicts more accurately than the benchmark in
a majority of the cases.
The alternative hypothesis $H_a$ is then obviously that the studied
model performs worse than the benchmark in a majority of the cases. 
The bootstrap algorithm
is as follows
\begin{enumerate}
\item Calculate the $\hat \epsilon$ values for the $n_w$ different sample windows
\item Pick $n_s$ samples of size $n_w$ from the $\hat \epsilon$ values
{\bf with replacement} 
\item Calculate the $\alpha/2$ and the $1-\alpha/2$ per cent points from the 
empirical distribution function of the selected samples, as the lower and upper 
confidence bounds respectively
\item Reject the null hypothesis and accept the alternative hypothesis if
    the upper bound is $< .5$, and accept the null hypothesis and reject the alternative
    hypothesis if the lower bound is $>.5$ at the 100$\alpha$ per cent significance level. 
    If the bound overlaps with $.5$ we say that the model performs {\it on par} with
    the benchmark.
\end{enumerate}
R code which implements this test is available in Appendix~\ref{sec:rcode}. 
This Monte Carlo bootstrap algorithm is used for two reasons, first to 
avoid making any assumptions about the distribution of the 
normalized MSEs in the test, and second to easily 
map MSE uncertainty to bounds on
our NDE statistic. The NDE bound $[lower,upper]$ 
can be interpreted as there
being a 100$(1-\alpha)$ per cent likelihood of the model performing better
than the random walk model in 100$\cdot lower$ per cent to
100$\cdot upper$ per cent of the cases.   

In the following sections we apply this evaluation method 
to the PlanetLab and Tycoon series.

\section{PlanetLab Analysis}\label{sec:pl}
PlanetLab (PL) is a {\it planetary-scale},
distributed computing platform comprising approximately 726 machines
at 354 sites in 25 countries, all running the same Linux based
operating system and PlanetLab software.
The user community is predominantly computer science
researchers performing large-scale networking algorithm and system
experiments. The time series is from December 2005 to December
2006. We calculate demand by aggregating the load value across all
hosts and averaging in hourly intervals with a 5-min sample 
granularity. This load measures
the number of processes that are ready to run 
on machine. 

\subsection{Model}
\begin{figure*} [htp]
  \begin{center}
    \subfigure[Sample Series]{\label{fig:pl_ts}\includegraphics[scale=0.4]{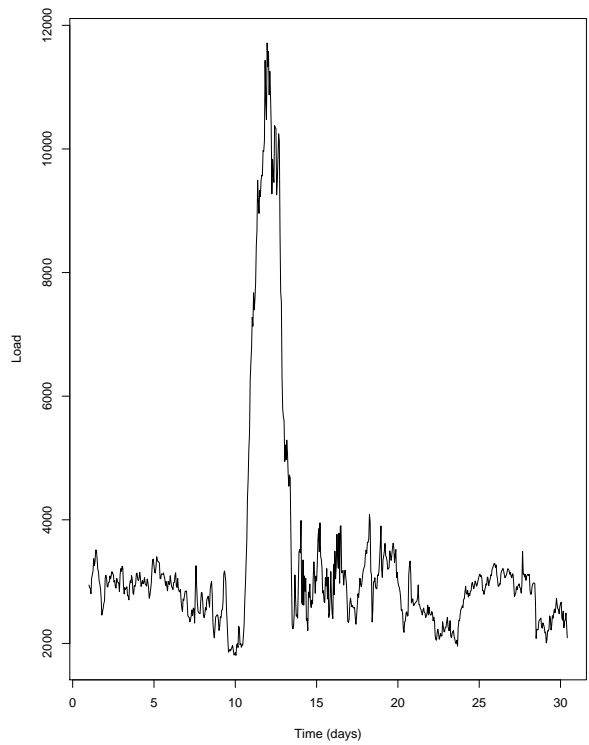}}
    \subfigure[Full Series]{\label{fig:pl_all_ts}\includegraphics[scale=0.4]{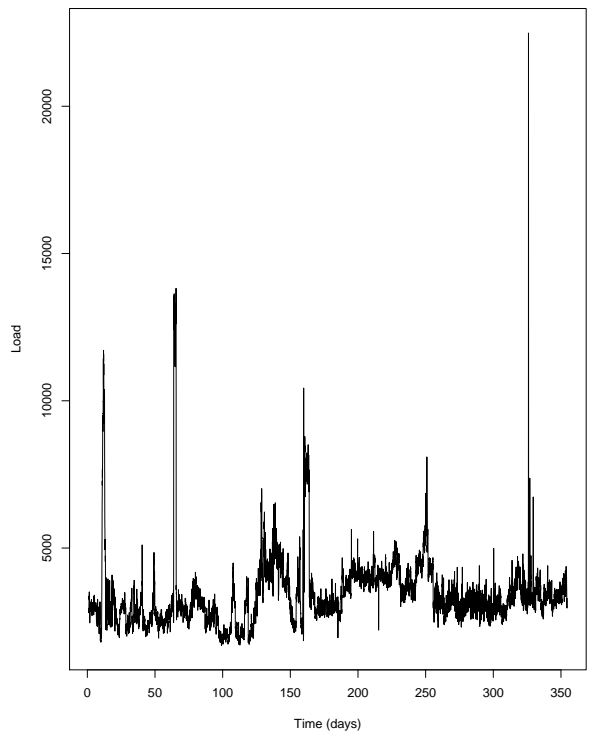}}
    \subfigure[Box-Cox Transform]{\label{fig:pl_boxcox}\includegraphics[scale=0.4]{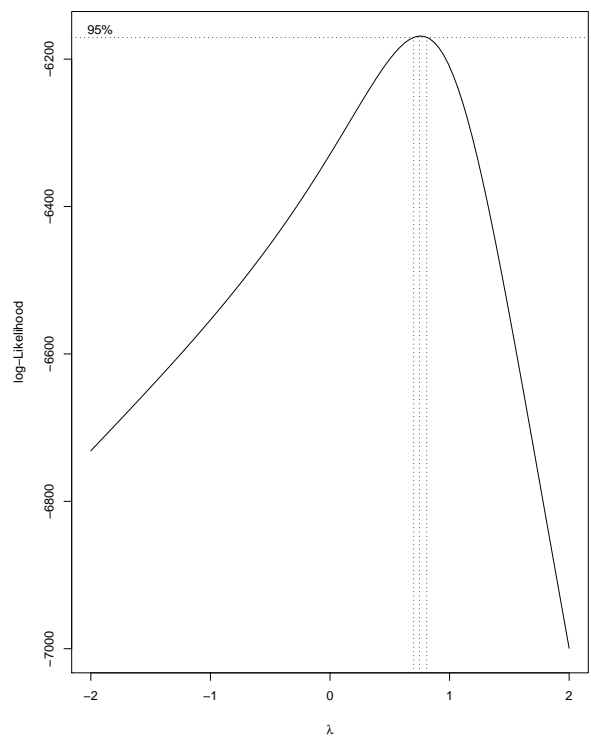}}
  \end{center}
  \caption{\small PlanetLab Series}
  \label{fig:pl_orig1}
\end{figure*}

\begin{figure*} [htp]
  \begin{center}
    \subfigure[Autocorrelation Function]{\label{fig:pl_acf}\includegraphics[scale=0.4]{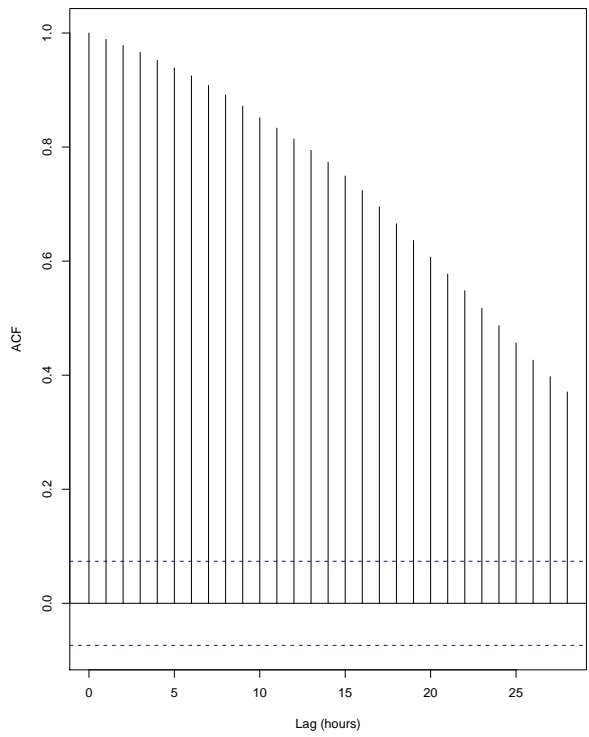}}
    \subfigure[Partial Autocorrelation Function]{\label{fig:pl_pacf}\includegraphics[scale=0.4]{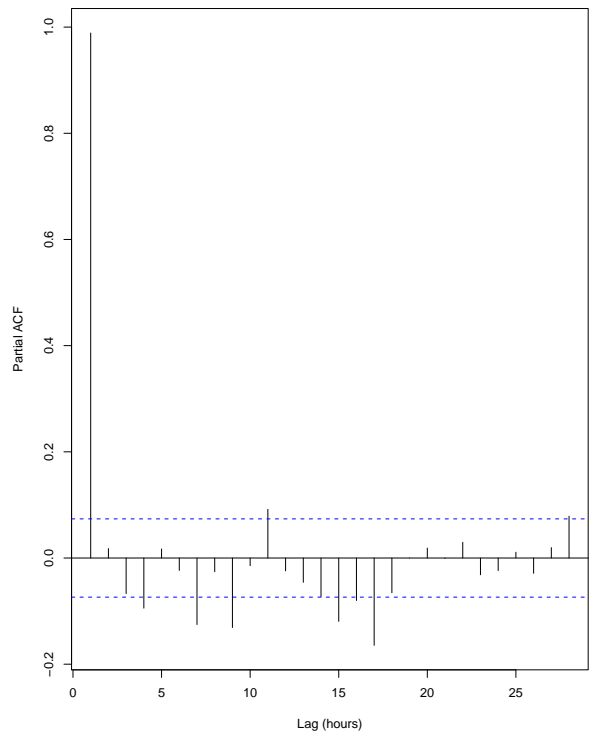}}
    \subfigure[Differenced Autocorrelation Function]{\label{fig:pl_diff_acf}\includegraphics[scale=0.4]{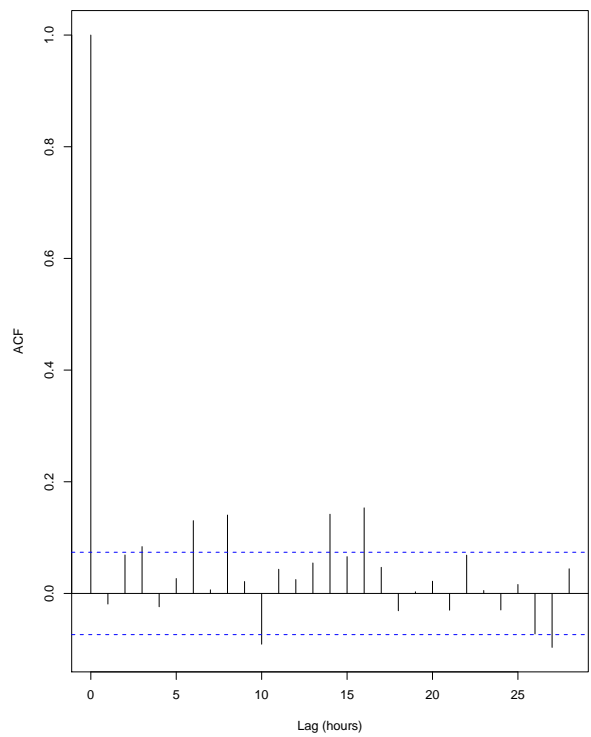}}
  \end{center}
  \caption{\small PlanetLab Autocorrelation Functions}
  \label{fig:pl_orig2}
\end{figure*}

We select the first month of the trace (707 values out of 8485) as our 
sample to construct the general ARIMA model. 
The sample series is shown in Figure~\ref{fig:pl_orig1}. 
There is one big spike in the sample, and we
might be tempted to treat it as an outlier, but as seen from the full 
trace these spikes are quite common and thus need to be accounted
for in our model. We instead perform a Box-Cox~\cite{box1964} transform
to address non-stationarity in variance. The
Box-Cox plot for the sample is shown in Figure~\ref{fig:pl_boxcox}.
A $\lambda$ value of $0.8$ is thus used to transform the series
prior to the ARIMA analysis. This $\lambda$ value is somewhere
between a $\sqrt Z_t$ and a $Z_t$ (no) transform. 
From Figure~\ref{fig:pl_orig2} we note that
the ACF has a slow decline in correlation, and that the PACF is
near unit root in lag 1. 
Now to formally test for unit root we perform the
augmented Dickey-Fuller test~\cite{dickey1981}, and obtain
a t-statistic of $-2.0294$ which has an absolute value less than
the $5$ per cent critical value $-3.41$, so we cannot reject the null
hypothesis of a unit root.

Therefore, we difference the series and then see
that the differenced ACF in Figure~\ref{fig:pl_diff_acf},
does not exhibit any clearly significant 
correlations. Hence, we model the series as
as an ARIMA(0,1,0) process or random walk. 
We note that there appears to be small significant
seasonal correlations around lags 6,8,10,14 and 16. But we 
decide to ignore
those because of our small sample size, and to keep the 
predictor simple.
To summarize, the entertained model is
\begin{equation}
(1-B)Z_t=a_t
\end{equation}
where $B$ is the backshift operator and $a_t$ is the residual white 
noise process. 
A Box-Ljung test~\cite{box1978} of serial correlations
of the residuals of this model gives a $\chi^2$ value of
$228.297$ and a p-value of $4.974 \cdot 10^{-12}$ for $100$ degrees of freedom, 
so there is still structure unaccounted for. Our
tests showed that at least an ARIMA(16,1,0) model was needed
before the Box-Ljung test succeeded, which is not practical for our
purposes, so we stick to our ARIMA(0,1,0) model.
Because this model is one of our standard benchmarks (RW) we also add 
an ARIMA(1,1,0) model to our evaluation to simplify comparison.

\subsection{Forecast}
\begin{figure*} [htp]
  \begin{center}
    \subfigure[one-step]{\label{fig:pl_mse_1}\includegraphics[scale=0.4]{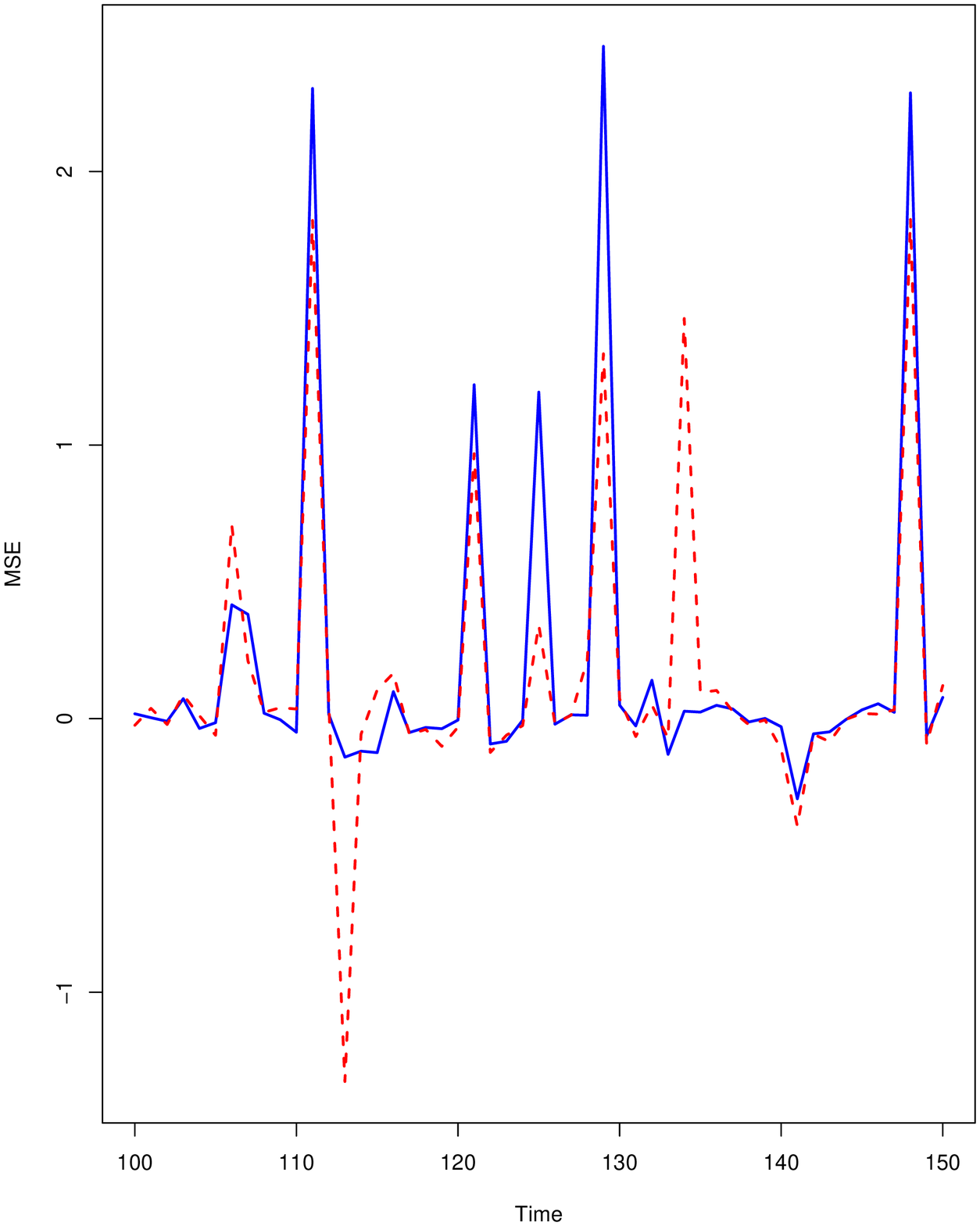}}
    \subfigure[two-step]{\label{fig:pl_mse_2}\includegraphics[scale=0.4]{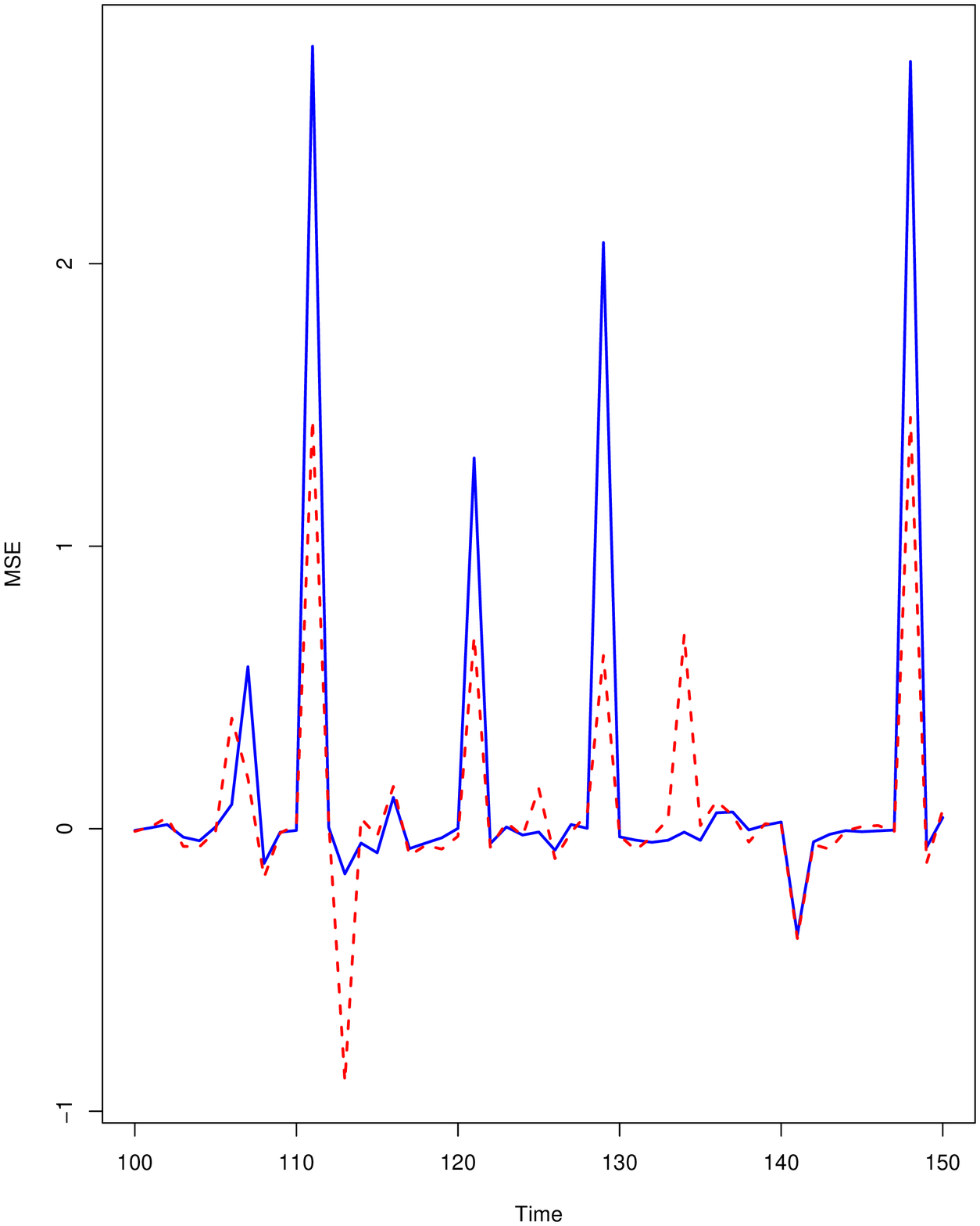}}
    \subfigure[three-step]{\label{fig:pl_mse_3}\includegraphics[scale=0.4]{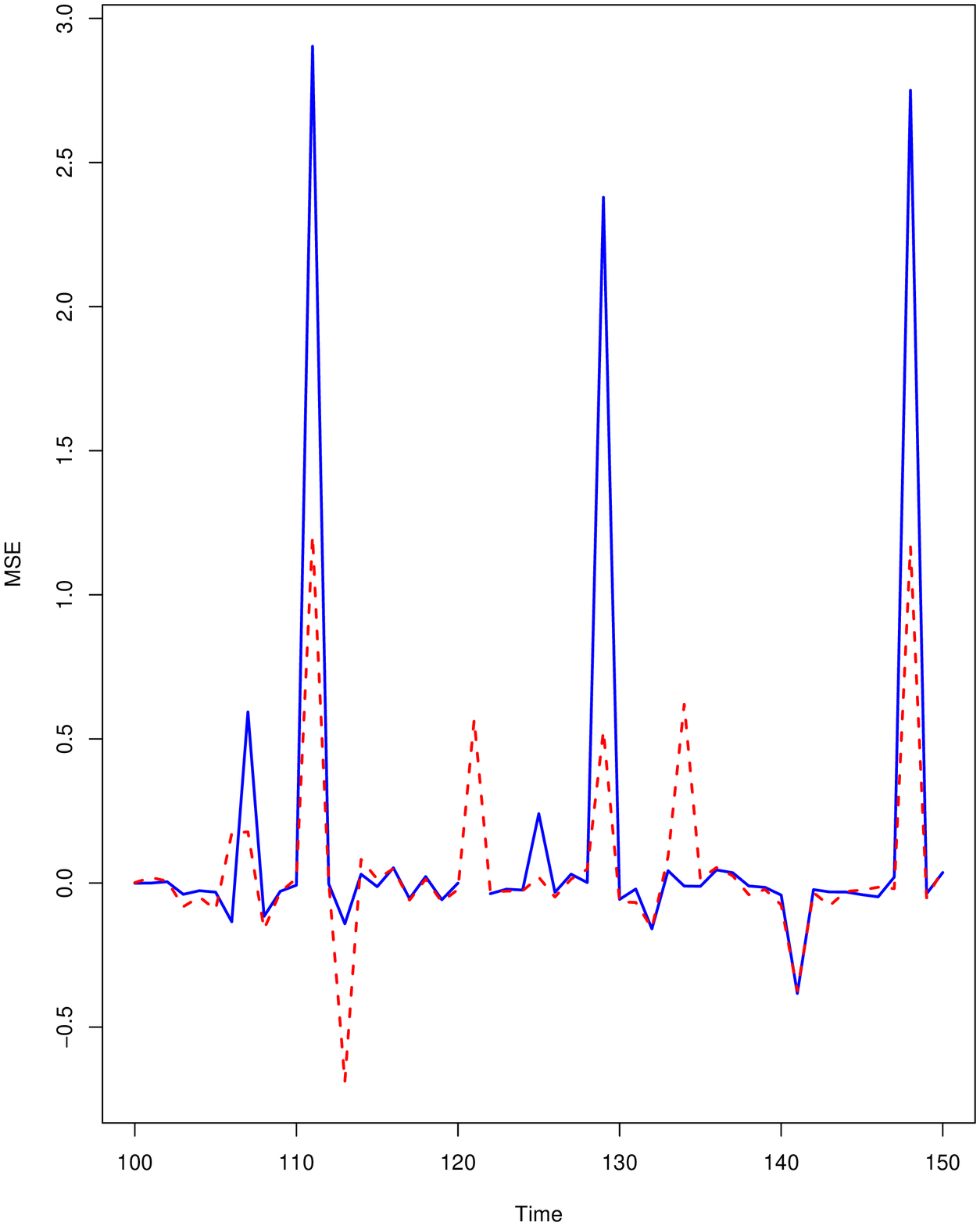}}
  \end{center}
  \caption{\small PlanetLab ARIMA(1,1,0) and Exponential Smoothing (dotted line) vs. Random Walk Model Forecast Errors}
  \label{fig:pl_mse}
\end{figure*}

We now compare the MSE of the one-, two- and three-step-ahead forecasts of the 
RW, EWMA, and ARIMA(1,1,0) models. The time windows used for predictions
range from $100$ to $150$ hours. Each empirical 
normalized MSE distribution thus has $50$ measurements.  
The evaluation of the forecasts of the 
ARIMA(1,1,0), and the exponential smoothing
models against the random walk model 
can be seen
in Figure~\ref{fig:pl_mse}. We note that a value less than $0$ in the plot
means that the model predictor performed better than the random walk predictor. 
We observe that both the ARIMA(1,1,0) and the
exponential smoothing model predictors seem to perform better than
the random walk predictor for the two and three-step ahead
predictions. 
We further note that there are more high peaks than deep valleys 
both for ARIMA(1,1,0) and EWMA, and that the EWMA peaks are lower.  
This pattern indicates that the RW model is more immune to extreme level 
shifts,
and that EWMA handles these shifts better than ARIMA(1,1,0).
  
Next, we use the statistical test constructed in the previous section
to verify the significance of the differences. 

\subsection{Statistical Test}
Table~\ref{T:pl_eval}
shows the NDE bound results for the PlanetLab models at significance 
level 5\% where 
$n_s$ was set to $1000$. The random walk row displays the errors in 
proportion to the true value observed, calculated as
\begin{equation}
\bar \epsilon = \frac{1}{T}\sum_{t=1}^T\frac{|\hat y_t - y_t|}{y_t} 
\end{equation}
where $\hat y_t$ is the predicted value at time $t$ and
$y_t$ is the actual value; and $T$ is the number of time windows
used in the test ($50$).
 We see that the errors 
ranged from $4.07$\% to $6.98$\% with the longer 
horizon forecasts performing worse.
From the NDE statistic bounds for 
the ARIMA(1,1,0) and EWMA rows in Table~\ref{T:pl_eval}
we can conclude that the ARIMA(1,1,0) model generates
predictions on par with the random walk model, for one and
two-step-ahead predictions, and better at significance level
$5$ per cent for three-step-ahead forecasts.
The EWMA model performs better for longer forecasts but not
at a significant enough level to pass our test.
To summarize, the only strong conclusion we can draw 
from these simulations
is that the ARIMA(1,1,0) predictor performed better than a random walk
predictor for three-hour ahead forecasts, but in general the RW model
selected performs relatively well. 

\begin{table}
\caption{\small PlanetLab Model NDE Bounds at $5$\% Significance Level with
                Random Walk (RW), Exponentional Smoothing (EWMA) and ARIMA(1,1,0) models,
                using 1,2 and 3-step ahead (SE) Forecasts.}
\label{T:pl_eval}
\begin{center}
\begin{tabular}{|l|c|c|c|}
  \hline
      & 1 SE & 2 SE & 3 SE \\
  \hline
  \hline
RW   &  $.0407$ & $.0554$ & $.0698$ \\  
  \hline
ARIMA(1,1,0) & $[.353,.627]$  & $[.471,.725]$ & $[.540,.800]$   \\  
  \hline
EWMA & $[.314,.588]$   & $[.373,.647]$   &  $[.392,.667]$ \\  
  \hline
\end{tabular}
\end{center}
\end{table}

\section{Tycoon Analysis}\label{sec:tycoon}
Tycoon is a computational market where resources, such
as CPU, disk, memory, and bandwidth can be purchased
on demand to construct ad-hoc virtual machines.
The price of the resources is in direct proportion to the 
demand, in that the cost of a resource share is dynamically
calculated as the ratio between the bid a user places on the 
resource and the bids all other users of that resource place.
The Tycoon network currently comprises about 70 hosts. Usage
is sparse and spiky, and is mostly generated from 
different test suites that are designed to evaluate the system. 
A trace was recorded of the aggregated CPU price 
in hourly intervals with a 10-min granularity during a period of 
17 days in July-August 2007. 

\subsection{Model}
\begin{figure*} [htp]
  \begin{center}
    \subfigure[Sample Series]{\label{fig:ty_ts}\includegraphics[scale=0.4]{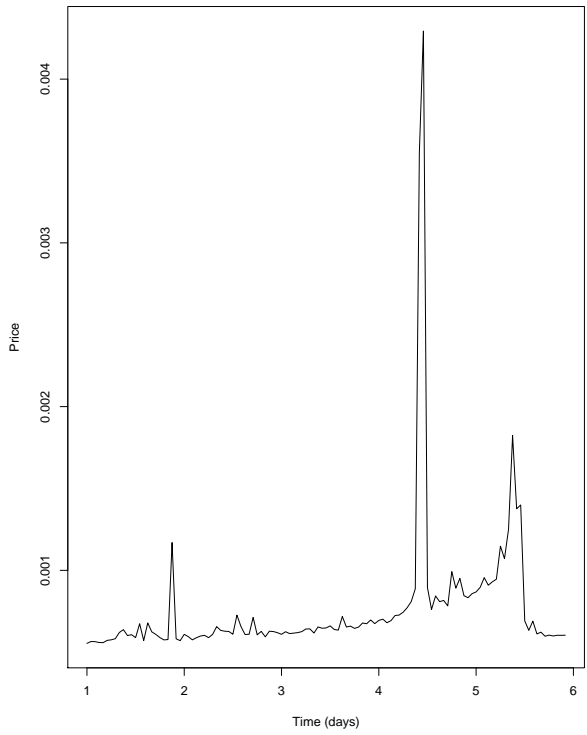}}
    \subfigure[Full Series]{\label{fig:ty_all_ts}\includegraphics[scale=0.4]{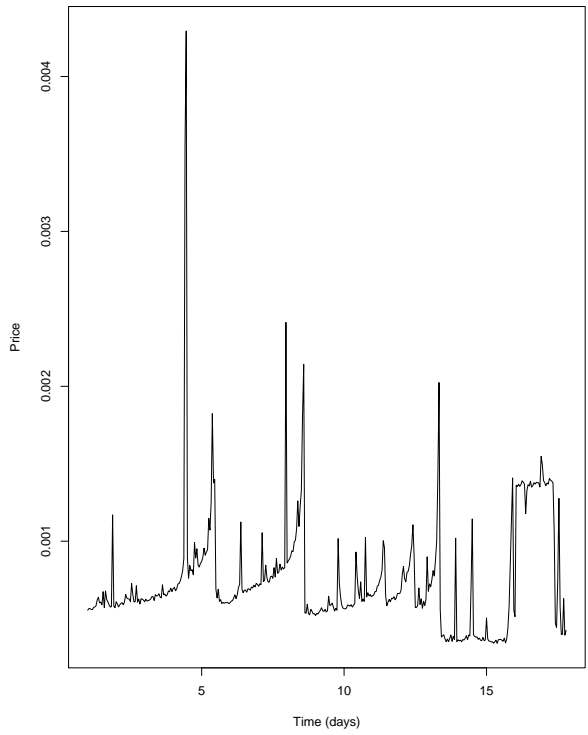}}
    \subfigure[Box-Cox Transform]{\label{fig:ty_boxcox}\includegraphics[scale=0.4]{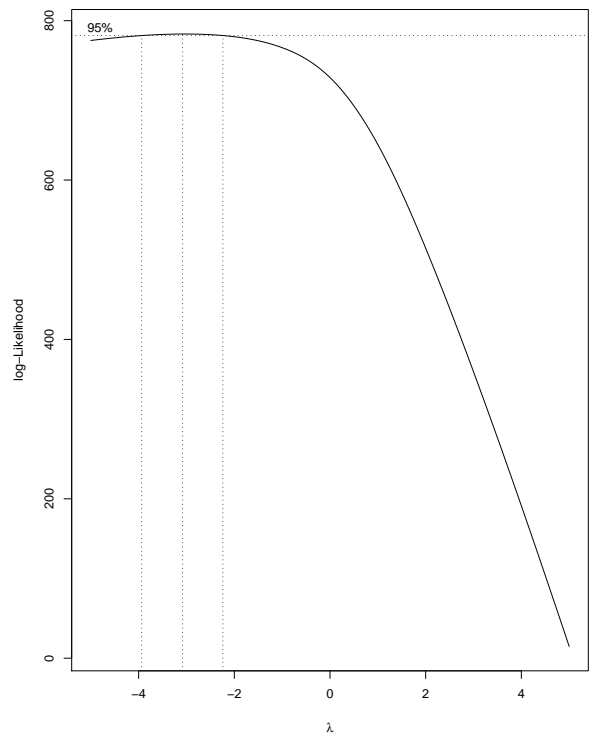}}
  \end{center}
  \caption{\small Tycoon Series}
  \label{fig:ty_orig1}
\end{figure*}

\begin{figure*} [htp]
  \begin{center}
    \subfigure[Autocorrelation Function]{\label{fig:ty_acf}\includegraphics[scale=0.4]{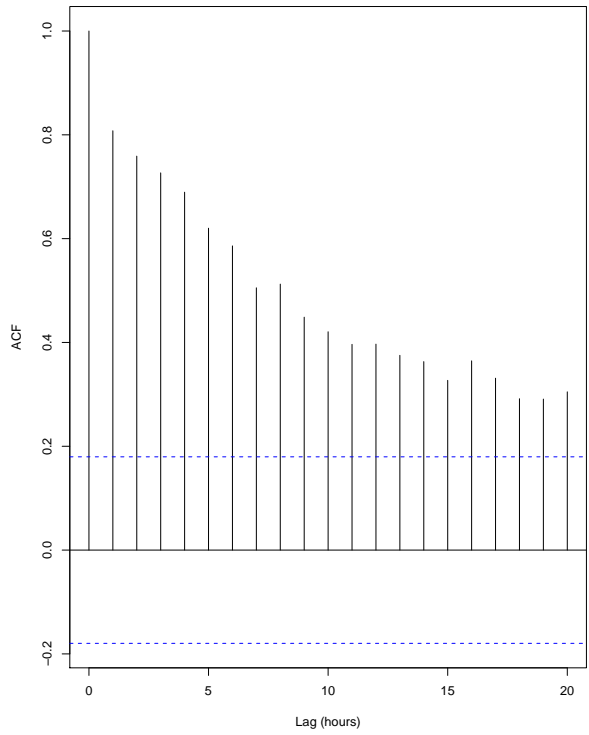}}
    \subfigure[Partial Autocorrelation Function]{\label{fig:ty_pacf}\includegraphics[scale=0.4]{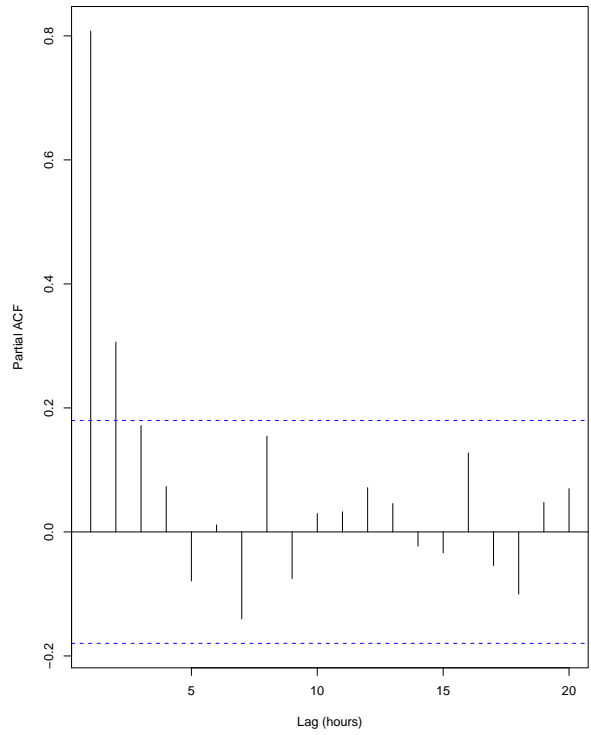}}
    \subfigure[Differenced Autocorrelation Function]{\label{fig:ty_diff_acf}\includegraphics[scale=0.4]{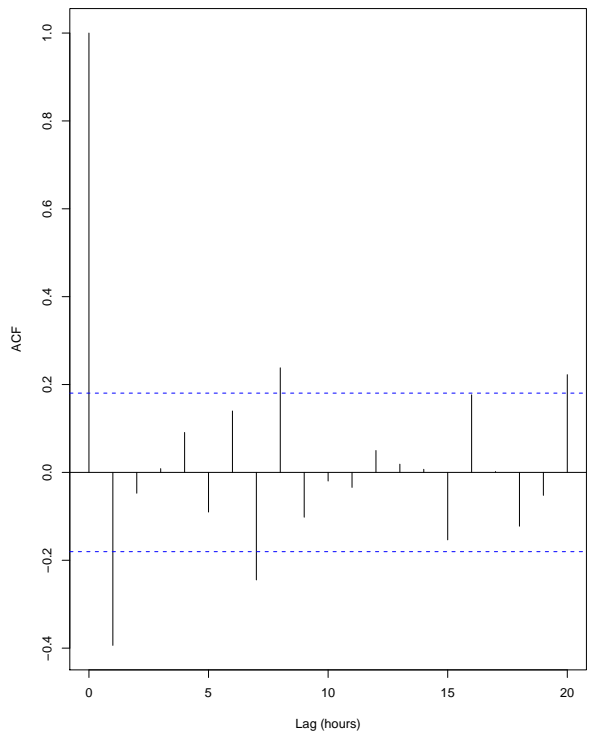}}
    \subfigure[ARIMA(1,1,0) Residuals Autocorrelation Function]{\label{fig:ty_res}\includegraphics[scale=0.4]{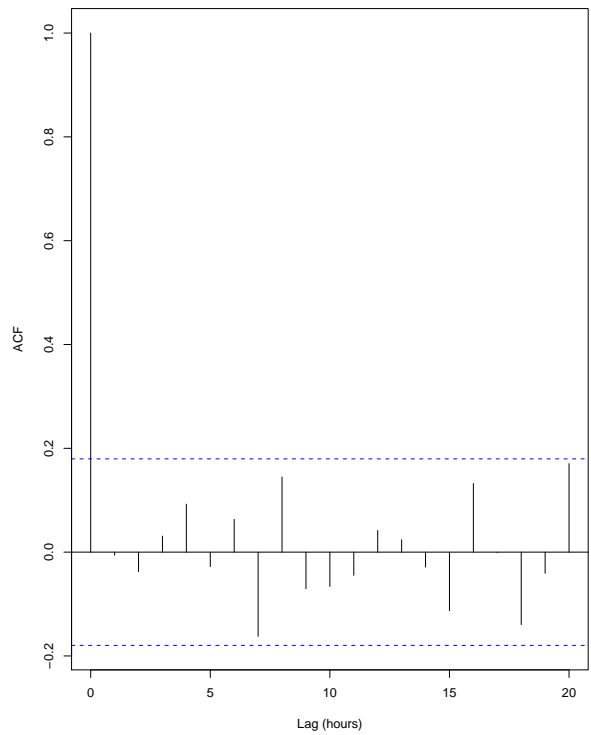}}
  \end{center}
  \caption{\small Tycoon Autocorrelation Functions}
  \label{fig:ty_orig2}
\end{figure*}

We select the first five days of the trace (119 values out of 404)
as our sample to
construct the general ARIMA model.  
The sample and the full series are shown in Figure~\ref{fig:ty_orig1}. 
Due to suspected non-stationarity in variance a Box-Cox
transform is again performed. As seen in Figure~\ref{fig:ty_boxcox},
the $\lambda$ value obtained was $-3$.
We note that
the ACF decays slowly and the PACF has a high first lag in Figure~\ref{fig:ty_orig2}.
So we again difference the series. Now the ACF shows only
one significant lag, so we can model it as an IMA(1,1) process.
The correlations of the residuals of this model
can be seen in Figure~\ref{fig:ty_res}. 
To summarize, the entertained model is
\begin{equation}
(1-B)Z_t=(1-\theta B)a_t
\end{equation}
where $B$ is the backshift operator and $a_t$ is the residual white noise process.
The $\theta$ coefficient was found to be statistically 
significant at a $5$ per cent significance level, and was fit to $.511$.
A Box-Ljung test of serial correlations
of the residuals of this model gives a $\chi^2$ value of
$87.758$ and a p-value of $.804$ for $100$ degrees of freedom, 
hence we conclude
that the model does not have any serial correlations
and is accurate. 
Because this model is one of our standard benchmarks (EWMA) we also add 
an ARIMA(1,1,0) model to our evaluation to simplify comparison.

\subsection{Forecast}
\begin{figure*} [htp]
  \begin{center}
    \subfigure[one-step]{\label{fig:ty_mse_1}\includegraphics[scale=0.4]{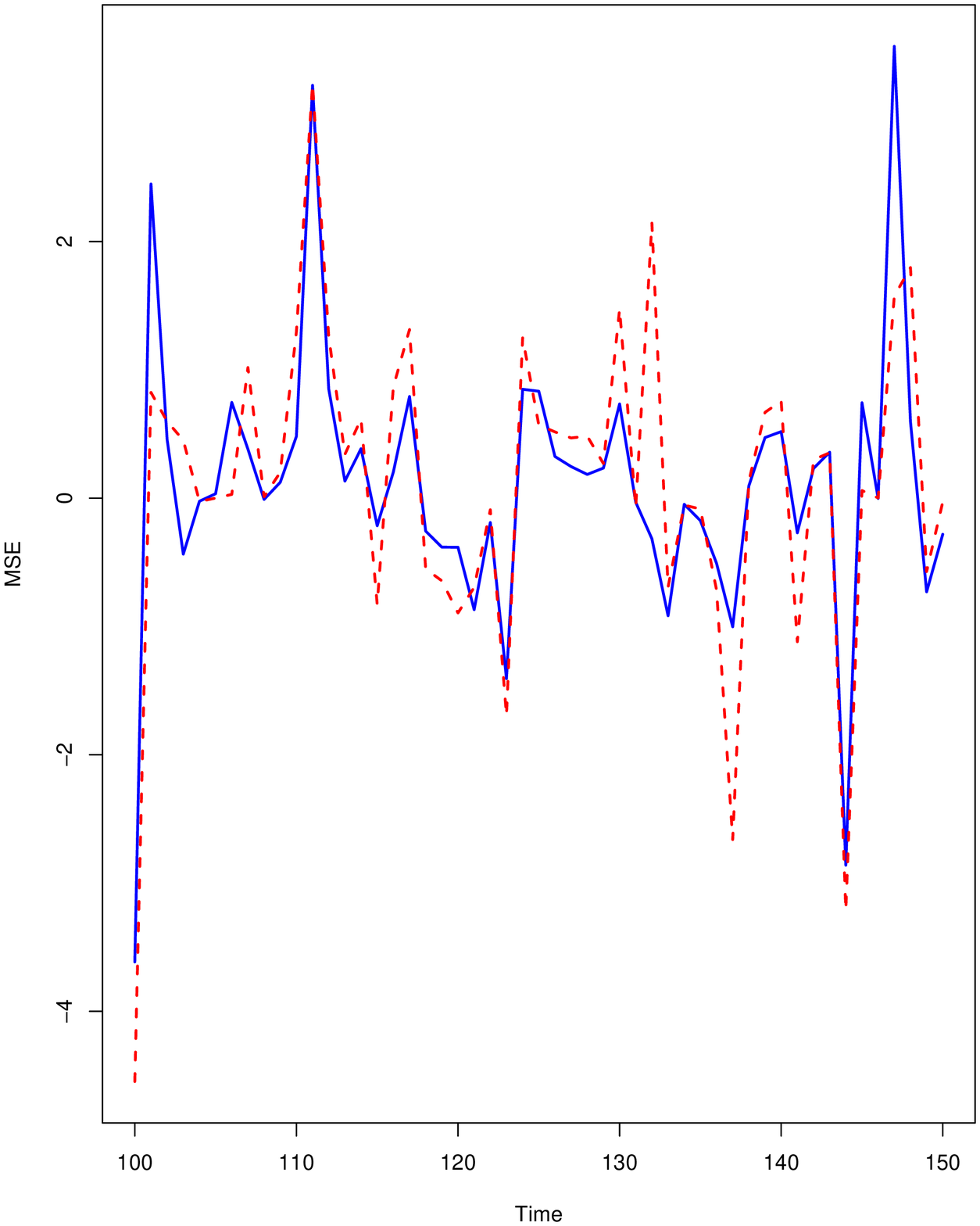}}
    \subfigure[two-step]{\label{fig:ty_mse_2}\includegraphics[scale=0.4]{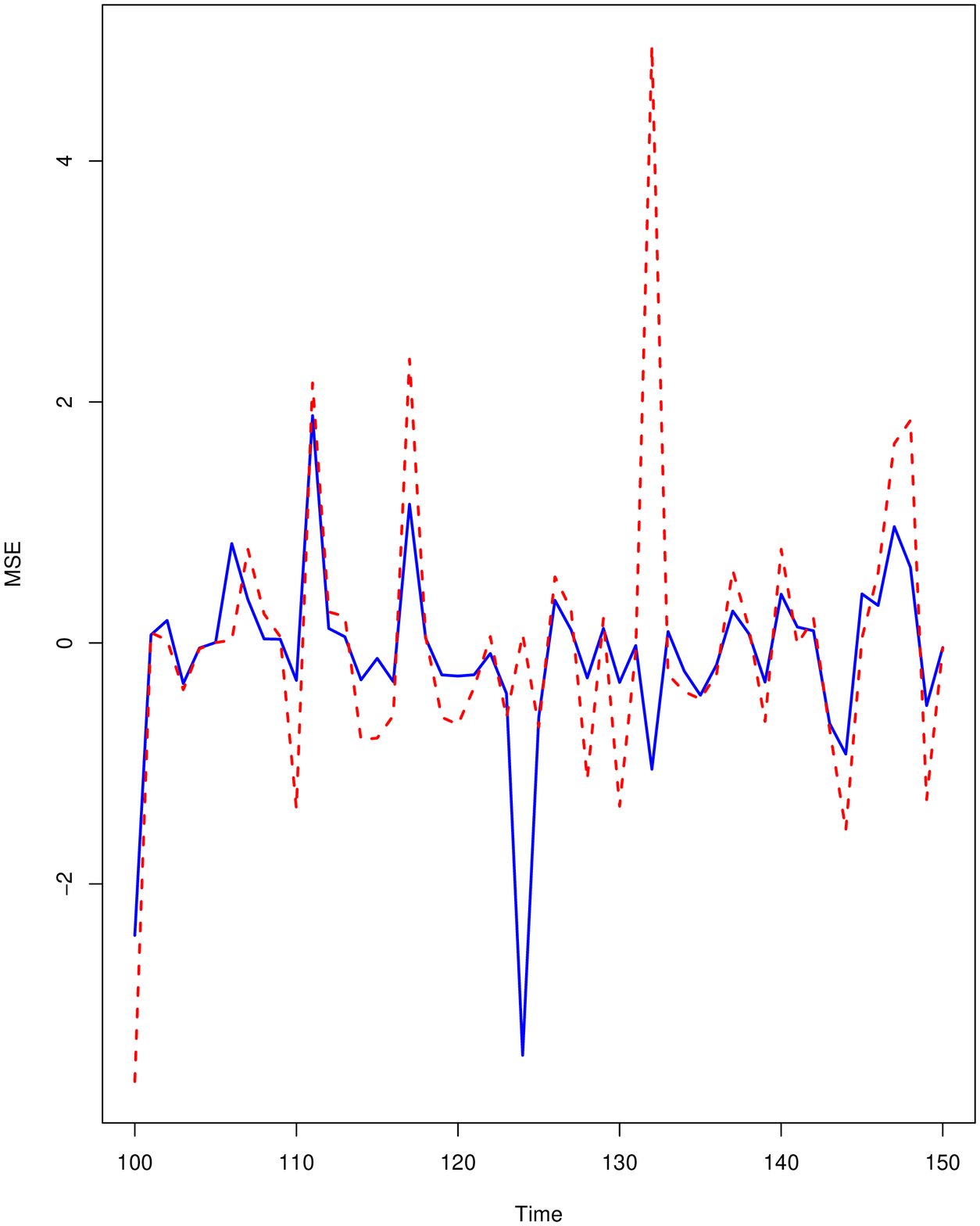}}
    \subfigure[three-step]{\label{fig:ty_mse_3}\includegraphics[scale=0.4]{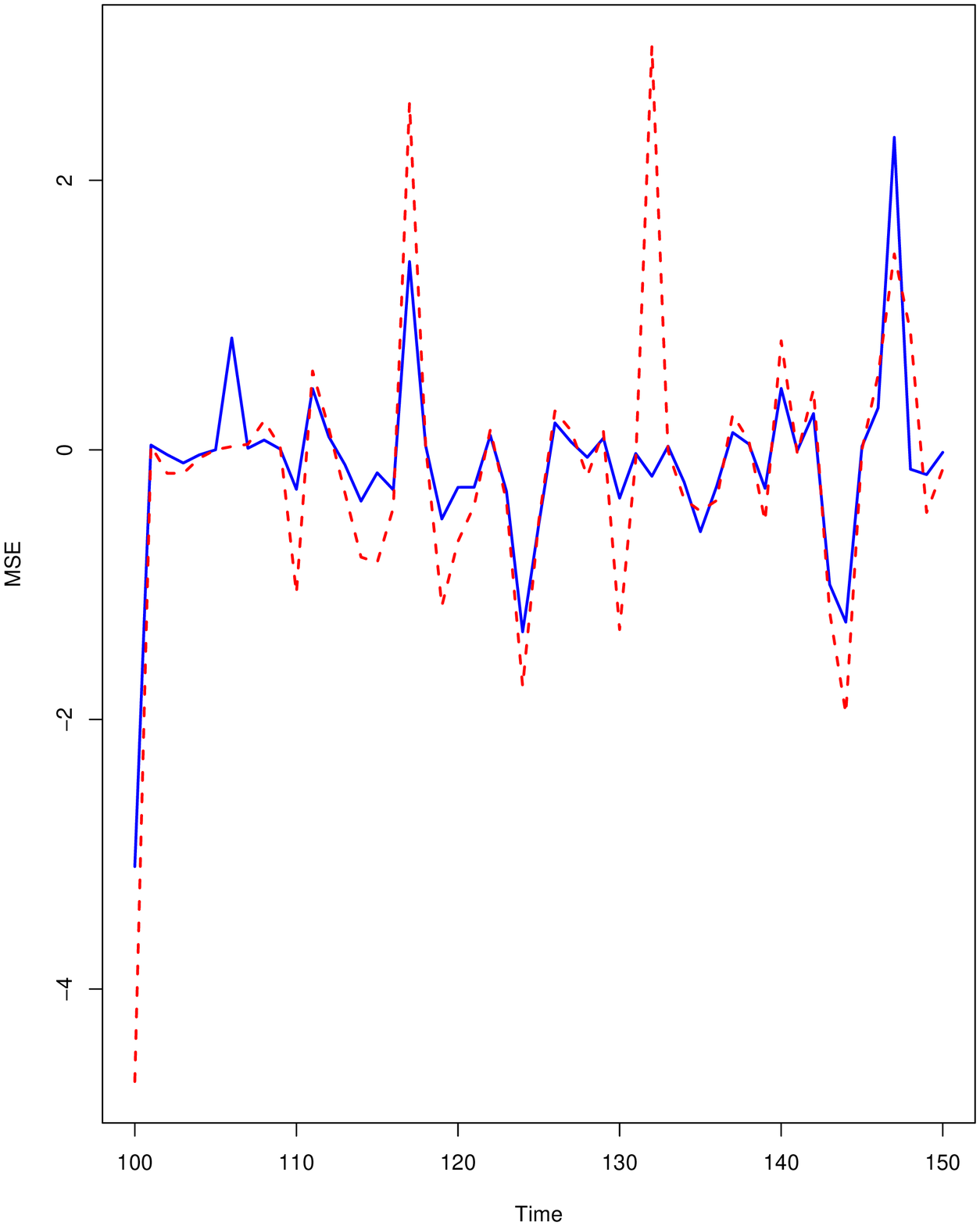}}
  \end{center}
  \caption{\small Tycoon IMA and Exponential Smoothing (dotted line) vs. Random Walk Model Forecast Errors}
  \label{fig:ty_mse}
\end{figure*}

We now compare the MSE of the forecasts of the RW,
EWMA, and ARIMA(1,1,0) models.
The model parameters are evaluated before each forecast. The time-window
used for model fitting ranged from 50 hours to 100 hours into the past, 
and thus again $50$ measurements were generated.
The evaluation of the ARIMA(1,1,0), and the exponential smoothing models
against the random walk model is shown in Figure~\ref{fig:ty_mse}. 
We recall that a value less than $0$ in the plot
means that the model predictor performed better than the random walk predictor.
It is not as clear as in the PlanetLab series that RW has fewer extremes
of bad predictions. However the ARIMA(1,1,0) model does seem to produce
less extreme peaks and valleys than EWMA, i.e. the 
opposite of what was observed for the PlanetLab data. 
Due to high volatility it is difficult to draw any conclusions about 
which model performs best from these plots, so we again have to resort to 
our statistical test.

\subsection{Statistical Test}
Table~\ref{T:ty_eval}
shows the NDE bound results for the Tycoon models at 
significance level 5 per cent where 
$n_s$ was set to $1000$. We see that the RW model performed much worse
for this time series compared to in the PlanetLab series. 
Average errors range from
$14.4$ per cent to $24.3$ per cent. This apparent difficulty in predicting the series
also reflects the results. We see that both the
ARIMA(1,1,0) and EWMA models performed on par with RW for 
all forecasts.
So at the $5$ per cent
significance level no strong conclusions can be drawn about which model 
performed best.
We however note, for the three step-ahead forecasts, that ARIMA(1,1,0) 
is close to being significantly better than RW, and for one step-ahead
forecast, EWMA is close to being significantly worse than RW.
The same pattern is apparent here, as in the PlanetLab data; 
the higher order ARIMA models
perform better for longer forecasts.
\begin{table}
\caption{\small Tycoon Model NDE Bounds at $5$\% Significance Level with
                Random Walk (RW) and Exponentional Smoothing (Exp) Benchmarks,
                using 1,2 and 3-step ahead Forecasts.}
\label{T:ty_eval}
\begin{center}
\begin{tabular}{|l|c|c|c|}
  \hline
    & 1 SE & 2 SE & 3 SE \\
  \hline
  \hline
RW & $.144$ & $.199$ & $.243$ \\
  \hline
ARIMA(1,1,0) & $[.333,.588]$  & $[.392,.647]$ & $[.431,.706]$    \\  
  \hline
EWMA     & $[.255,.510]$ & $[.353,.627]$ & $[.412,.686]$   \\  
  \hline
\end{tabular}
\end{center}
\end{table}

\section{Series Comparison}\label{sec:compare}
In this section we compare the dynamics of
the PlanetLab series to the Tycoon series using the
full traces, and give
both quantitative and qualitative explanations to
the differences.

Table~\ref{T:quart} shows the range and the quartiles of the series,
normalized by the series mean. The Tycoon series has a median which is 
further away from the mean, and the range of values is slightly
tighter. The narrower range is expected because of the
time horizon difference in the two series. 
However, overall the statistics for PlanetLab and Tycoon are
strikingly similar. This is a bit surprising since 
Tycoon is just in an early test phase with very limited usage and demand,
whereas PlanetLab is a mature system that has been in operation for
several years.
\begin{table}
\caption{\small Mean Normalized Quartiles and Range}
\label{T:quart}
\begin{center}
\begin{tabular}{|l|c|c|c|c|c|c|}
  \hline
 & Min & Q1 & Median & Q3 & Max \\
  \hline
  \hline
PlanetLab & .494 & .811 & .936 & 1.12 & 6.55  \\
  \hline
Tycoon & .452 & .763 & .860 & 1.10 & 5.70  \\
  \hline
\end{tabular}
\end{center}
\end{table}

\begin{figure*} [htp]
  \begin{center}
    \subfigure[PlanetLab Residual Autocorrelation Function]{\label{fig:pl_noise}\includegraphics[scale=0.4]{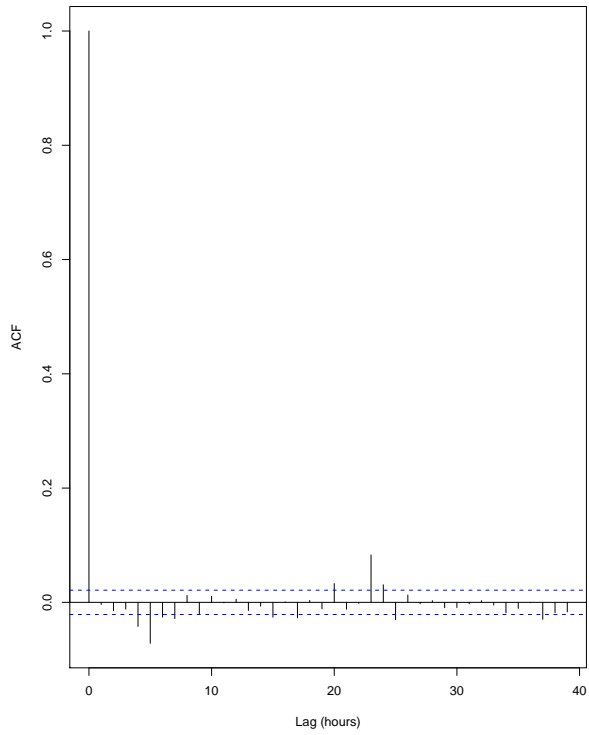}}
    \subfigure[PlanetLab Squared Residual Partial Autocorrelation Function]{\label{fig:pl_noise_sqr}\includegraphics[scale=0.4]{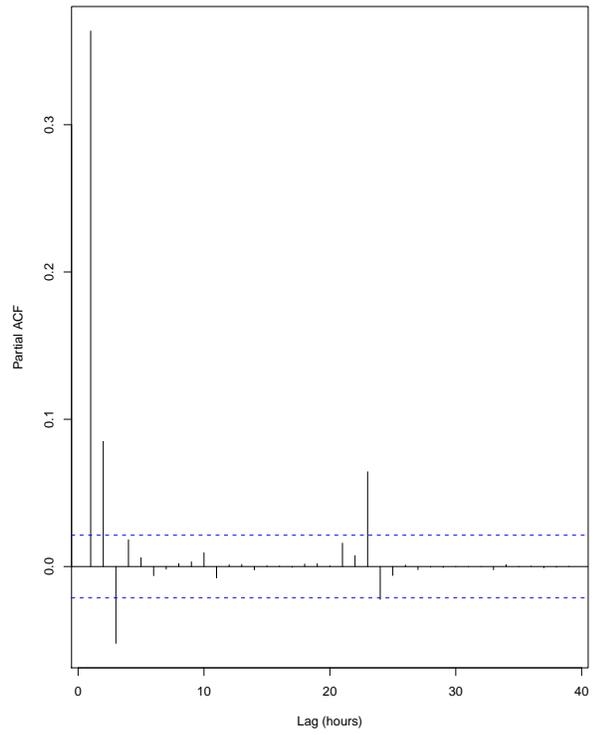}}
    \subfigure[Tycoon Residual Autocorrelation Function]{\label{fig:ty_noise}\includegraphics[scale=0.4]{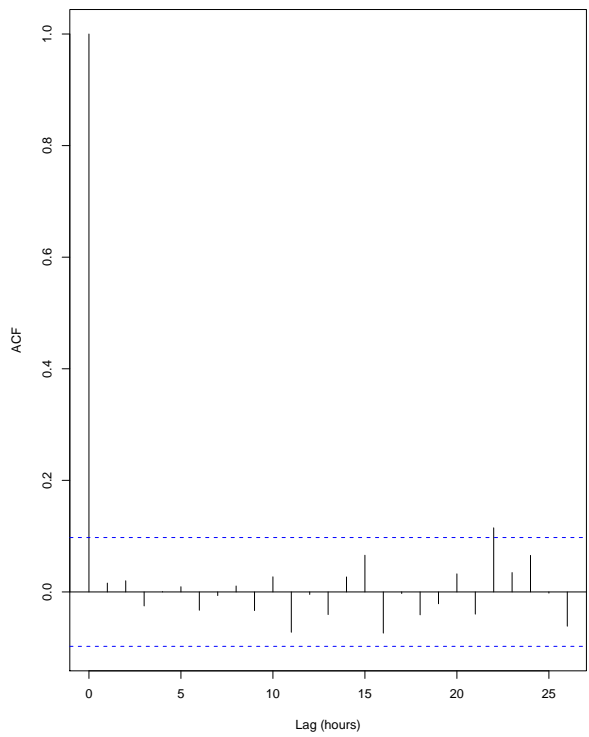}}
    \subfigure[Tycoon Squared Residual Partial Autocorrelation Function]{\label{fig:ty_noise_sqr}\includegraphics[scale=0.4]{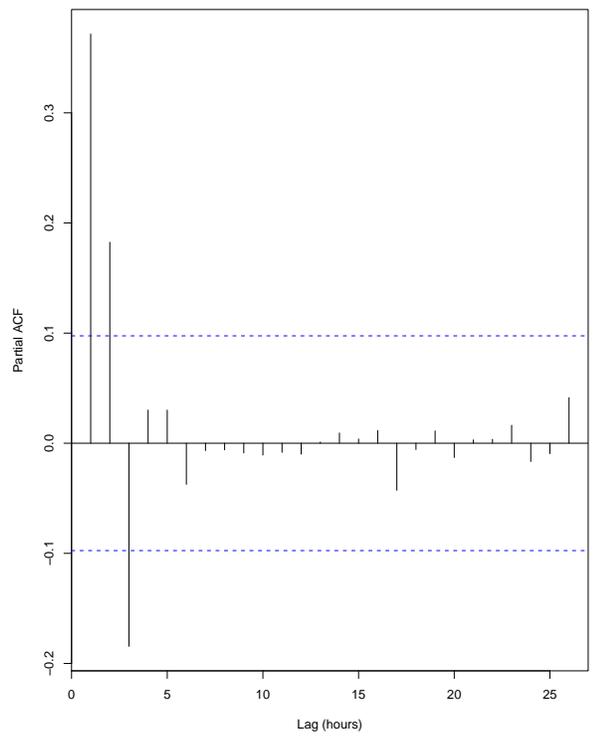}}
  \end{center}
  \caption{\small PlanetLab and Tycoon Volatility Analysis}
  \label{fig:vol}
\end{figure*}

The volatility statistics of the two series are compared in
Table~\ref{T:vol}. We conclude that the variance is higher
in Tycoon, but the right tail of the PlanetLab series
distribution is longer, and the PlanetLab series is also 
more prone to outliers. Again it is remarkable how closely
the tail and outlier behavior of the much smaller 
Tycoon sample follows the PlanetLab statistics.
\begin{table}
\caption{\small Volatility Characteristics}
\label{T:vol}
\begin{center}
\begin{tabular}{|l|c|c|c|}
  \hline
 & Coef of Variation & Skewness & Kurtosis \\
  \hline
  \hline
PlanetLab & .362 & 4.03 & 28.29 \\
  \hline
Tycoon & .511 & 3.67 & 24.38 \\
  \hline
\end{tabular}
\end{center}
\end{table}
To determine whether any of these series exhibit heteroskedasticity,
we take the squared residuals from an ARIMA model of the full
series and fit an AR model.
Then according to Engle~\cite{engle1982} heteroskedasticity exists if
\begin{equation}
    1-\chi^2_s((n-s)R^2) < \alpha 
\end{equation}
where $n$ is the number of values in the series, $s$ the order of the AR model
fit to the squared residuals, $\chi^2_{df}$ is the $\chi^2$ density function
with $df$ degrees of freedom, and $\alpha$ is the significance level. 
The complete PlanetLab series follows an ARIMA(3,1,0) model and the
complete Tycoon series follows an IMA(1,2) model. The residuals and their
squares of these models can be seen in Figure~\ref{fig:vol}.

We find
that both the PlanetLab and Tycoon series pass the significance test 
at the $5$ per cent significance level. 
Furthermore, both the Tycoon and the PlanetLab squared residuals  
follow AR(3) models, i.e., they have very similar volatility
dynamics structure.

It is easy to see that this heteroskedasticity could cause more outliers
and higher kurtosis in a static model. Intuitively, if the first moment
fluctuates, the second moment increases, and similarly if the second moment fluctuates
there is a greater likelihood of more spikes or AR model outliers, which
would increase the kurtosis. High volatility 
and dynamics in structure could also explain why ARIMA predictions assuming
static volatility and regression structure perform so poorly compared to 
a simple random walk predictor. However, we note that a random walk predictor
does not accurately estimate risk of high demand, which is more apparent for forecasts
with a longer future time horizon. An alternative approach to studying volatility and risk over time
is the approach of measuring long term memory or dependence. This was done in~\cite{sandholm2007b}, and
we found that non-Gaussian long term dependencies did exist, which could cause so called
workload flurries with abnormally high demand.  

The ARIMA(1,1,0) model performs better 
in PlanetLab 
than in Tycoon, which may indicate that PlanetLab has longer memory of 
past values than Tycoon. This may be attributed to the shorter sample period
and the nature of the applications currently running on Tycoon; mostly
short intense test applications.

\section{Related Work}\label{sec:related_work}
The algorithm used for the statistical test
of significant differences in predictor
performance was inspired by the 
Monte Carlo bootstrap method introduced
by Efron in~\cite{efron1979} and popularized
by Diaconis and Efron in~\cite{diaconis1983}. The bootstrap
method is typically used as a non-parametric
approach to making confidence claims.
We, use it to expand a short sample into a bigger one
without any distributional assumptions about the MSE terms. 
A more typical
usage is to shrink a large sample into multiple 
smaller random samples that are easier to make
statistical claims about collectively. 

Tycoon usage has not been statistically investigated
before. Previous work on the computational
market characteristics of Tycoon has used PlanetLab
and other super computing center job traces as a proxy for
expected market demand~\cite{sandholm2007a} or made 
simple Gaussian distribition, and Poisson arrival 
process assumptions~\cite{sandholm2007b}.

In this work we support the study of PlanetLab 
as a proxy for Tycoon demand, by verifying a large
number of statistical commonalities, both in terms
of structure of series and in terms of optimal predictor
strategies. Chun and Vahdat~\cite{chun2003}
have analyzed 
PlanetLab usage data but not from a predictability 
viewpoint. Their results 
include observations of highly bursty and order of magnitude
differences in utilization over time,
which we also provide evidence for. We note that the PlanetLab
trace that Chun and Vahdat
studied was from 2003.

Oppenheimer et al.~\cite{oppenheimer2006} also analyze 
PlanetLab resource usage and further evaluate
usage predictors and conclude
that mean reverting processes such as exponential
smoothing, median, adaptive median, 
sliding window average, adaptive average and running
average all perform worse than simple random
walk predictors and, what they call, {\it tendency predictors}
which assume that the trend in the recent past 
continues into the near future. They further notice no
seasonal correlations over time due to PlanetLab's 
global deployment. We do see some seasonal correlations
in our initial time series analysis but not significant enough
to take advantage of in predictions. Further, our evaluation approach
follows the traditional ARIMA model evaluation method, and
we provide a statistical test to verify and compare prediction efficiency.
One major difference between our studies and thus also the conclusions
is that Oppenheimer et al. only considered one-step ahead
predictions whereas we also consider two, and three-step ahead
predictors to do justice to the models considering correlations
beyond the last observed step. We finally note that they
studied PlanetLab data from August 2004 to January
2005, whereas we studied more recent data from
December 2005 to December 2006.

\section{Conclusions}\label{sec:conclusion}
This work set out to study the
predictive power of regression models
in shared computational networks such as
PlanetLab and Tycoon. The main result is
that no significant 
evidence was found that higher order regression
models performed better than
random walk predictions. The exception was for
three-step ahead predictions in PlanetLab
where an ARIMA(1,1,0) model outperformed 
the random walk model.

The study also shows the difficulty in composing
a model from a sample and then using this model
in predictions if the structure of the series
is changing over time as in the Tycoon case.

Our study highlighted
a number of statistical similarities
between Tycoon and PlanetLab,  such as
volatility structure, outlier likelihood,
and heavy right tails of density functions,
which motivates further studies and 
comparisons of workloads to improve
forecasting. 

The ARIMA models were refitted for every 50
to 150 hours to provide as accurate
models of the recent past as possible, but the
overall structure of the model was fixed as the
one obtained from the fit of the sample series.
Larger fitting windows were tested for the PlanetLab data
without any effect in the results, but larger windows could not
be tested for the Tycoon series due to the limited trace
time frame (17 days). 
There was however a clear pattern
that the higher order ARIMA models performed
better in the two and three-step ahead 
forecasts compared to the random walk model.

To summarize, we have exemplified the difficulties 
in modeling significant regressional parameters
for computational demand dynamics, even if the model is
very generic and the model parameters are re-estimated frequently. 
It was found difficult to improve on the random walk process
model for one-step-ahead forecasts, which is a bit surprising 
(and contradictory to the main hypothesis in~\cite{mandelbrot2004}) given
that RW processes, in theory, should generate a normal distribution of demand
whereas the actual measured demand distribution was very right skewed and 
heavy tailed, both in the PlanetLab and the Tycoon series.

We do however see that higher order regressional parameters can improve
the two-step and three-step ahead forecasts. More work is needed to determine
how these models should be discovered and dynamically updated. One possible 
extension is to see if there is an improvement in predictor performance if 
the model is allowed to changed dynamically as well as the parameters based on
observed ACF and PACF behavior. More work is also needed to determine the 
computational overhead of the more complicated regressional models and 
the calculations of fits and predictions. Accurate random walk
predictors can be built very easily with virtually no overhead,
so the improvement in accuracy needs to be significant to be worthwhile. 
This work does however show that there is a potential for improvement
of longer forecasts.

\section*{Acknowledgments}
I would like to thank Professor Magnus Boman for his detailed comments on
earlier versions of this paper; and Professor Raja Velu and Kevin Lai
for providing the inspirational ideas underlying this study.  

\bibliographystyle{latex8}
\bibliography{stat.bib}

\appendix
\section{Bootstrap Test R-Code}\label{sec:rcode}
\begin{mytinylisting}
\begin{verbatim}
predict_arima <- function(x,ord, window, horizon, mse, lambda) {
    n = (length(x)-window-2)/window
    errors=c()
    for (i in 0:n) {
        start_index = i * window
        stop_index = start_index + window -1
        outcome_index = start_index + window
        model = arima(boxcox_transform(x[start_index:stop_index],lambda), \
                               order=ord,method="ML")
        pred = predict(model, n.ahead=horizon)
        if (mse) {
            errors = c(errors, (x[outcome_index+horizon-1] - \
                                 boxcox_inverse(pred$pred[horizon],lambda))^2)
        } else {
            errors = c(errors, abs((x[outcome_index+horizon-1] - \
    boxcox_inverse(pred$pred[horizon],lambda))/x[outcome_index+horizon-1]))
        }
    }
    mean(errors)
}
evaluate_arima <- function(x,ord,from,stop,step,walk,horizon,mse,lambda) {
arima_mse = c()
walk_ind = 1
to = round((stop - from)/step) + from
  for (i in from:to) {
    window = from + ((i-from)*step)
    pred = predict_arima(x,ord,window,horizon,mse,lambda)
    if (length(walk) > 0) {
        pred = pred / walk[walk_ind]
        walk_ind = walk_ind + 1
    } 
    arima_mse=c(arima_mse,pred)
  }
  arima_mse
}
bootstrap_test <- function(x,sample_size,alpha) {
  n=length(x)
  x_sample = c()
  for (i in 1:sample_size) {
    x_sample=c(x_sample,ecdf(sample(x,n,replace=T))(0))
  }
  sort_sample = sort(x_sample)
  c( sort_sample[round(sample_size*alpha/2)], \
     sort_sample[round(sample_size*(1-alpha/2))] )
}
evaluate_walk_exp <- function(x,ord,horizons,from,to,step,alpha,samples,lambda)
{
  exp_errors = c()
  arima_errors = c()
  walk_errors = c()
  x_evals = c()
  x_exps = c()
  for (i in 1:horizons) {
    walk_error = evaluate_arima(x,c(0,1,0),from,to,step,c(),i,mse=F,lambda)
    walk_errors = c(walk_errors, mean(walk_error), mean(walk_error))
    x_walk = evaluate_arima(x,c(0,1,0),from,to,step,c(),i,mse=T,lambda)
    x_eval = evaluate_arima(x,ord,from,to,step,x_walk,i,mse=T,lambda)
    x_exp = evaluate_arima(x,c(0,1,1),from,to,step,x_walk,i,mse=T,lambda)
    x_evals = cbind(x_evals, x_eval)
    x_exps = cbind(x_exps, x_exp)
    # Pr(EXP < RW)
    exp_errors = c(exp_errors,bootstrap_test(log(x_exp),samples,alpha))
    # Pr(ARIMA < RW)
    arima_errors = c(arima_errors,bootstrap_test(log(x_eval),samples,alpha))
  }  
  errors = cbind(walk_errors,arima_errors,exp_errors)
  attr(errors,'arima') = x_evals
  attr(errors,'exp') = x_exps
  errors
}
\end{verbatim}
\end{mytinylisting}

\end{document}